\documentclass[11pt,fleqn]{article} %
\usepackage{psfig}
\title{Bifurcating Transport of Glassy Matter \\ Within Annular Micropores}  %
\author{Zotin K.-H.  Chu} 
\date{3/F, 4, Alley 2, Road Xiushan, Leshanxinchun, Xujiahui 200030, China}
\begin{document}
\maketitle
\begin{abstract}
Glassy matter, as subjected to high shear rates, exhibit shear
thinning : i.e., the viscosity diminishes with increasing shear
rate. 
Meanwhile one prominent
difference between the transport in  micropores and that in
macroscale  is the (relatively) larger roughness observed inside
micropores. As the pore size decreases, the surface-to-volume
ratio increases and therefore,  surface roughness will greatly
affect the transport in micropores. By treating the glass as a
shear-thinning matter and using the rate-dependent model together
with the boundary perturbation method, we can analytically obtain
the transport results up to the second order.
\newline

\newline
\end{abstract}
\doublerulesep=7mm        
\baselineskip=7mm \oddsidemargin-1mm
\bibliographystyle{plain}
\section{Introduction}
In recent years, considerable effort was geared towards
understanding how glasses respond to shear [1-3]. Phenomena such
as shear thinning  (i.e., the higher the shear rate is, the
smaller is the flow resistance) and 'rejuvenation' are common when
shear flow is imposed. Unlike crystals, glasses {\it age}, meaning
that their state depends on their history. When a glass falls out
of equilibrium, it evolves over very long time scales
[4-7].\newline Meanwhile most of the classical solutions of
contact problems, starting from the Hertzian case, rely on the
assumption of nominally smooth geometries, which is reasonable at
large enough scales. However, real surfaces are rough at the
micro- or even at the meso-scale [8-10]. 
The role of surface roughness has
been extensively investigated, and opposite conclusions have been
reached so far. For instance, friction can increase when two
opposing surfaces are made smoother (this is the case of cold
welding of highly polished metals). On the other hand, friction
increases with roughness when interlocking effects among the
asperities come into play. This apparent contradiction is due to
the effects of length scales, which appear to be of crucial
importance in this phenomenon.
\newline
In this paper we shall consider the transport of glassy matter
 in micropores which have radius- or
transverse-corrugations along the cross-section. The glassy matter
will be treated as a shear-shinning material. To consider the
transport of this kind of glass (shear-thinning fluids) in
microscopic domain, we adopt the verified transition-rate
dependent model [4-5] which was used to study the annealing of
glass.
\newline We noticed that Cagle and Eyring [5] tried a hyperbolic sine law
between the shear (strain) rate : $\dot{\xi}$ and (large) shear
stress : $\tau$ (because the relaxation at the beginning was
steeper than could be explained by the bimolecular law) and
obtained the close agreement with experimental data. They can
obtain the law of annealing of glass for explaining the too rapid
annealing at the earliest time.  This model has sound physical
foundation from the thermal activation process [4] (a kind of
(quantum) tunneling which relates to the matter rearranging by
surmounting a potential energy barrier was proposed therein; cf.,
Fig. 1).
With this model we can associate the (glassy) fluid with the
momentum transfer between neighboring atomic clusters on the
microscopic scale and reveals the atomic interaction in the
relaxation of flow with (viscous) dissipation. \newline The
outline of this short paper is as follows. Section 2 describes the
general mathematical and physical formulations of the framework.
In this Section, explicit derivations  for the glassy flow  are
introduced based on a  microscopic model proposed before [4]. The
boundary perturbation technique [8-9] will be implemented, too. In
the 3rd. and 4th. Sections, relevant results and discussion are
given therein.
\section{Mathematical and physical formulations}
By the beginning of this century the concept
of activation entropy was included in the model, and it was
considered that molecules go both in the forward direction
(product state) and in the backward direction (reactant state).
 The development of statistical mechanics, and later quantum
mechanics, led to the concept of the potential energy surface.
This was a very important step in our modem understanding of
atomic models of deformation. Eyring's contribution to this
subject was the formal development of the transition state theory
which provided the basis for deformation kinetics, as well as all
other thermally activated processes, such as crystallisation,
diffusion, polymerisation. etc. [4-5]\newline
The motion of atoms is represented in the configuration space; on
the potential surface the stable molecules are in the valleys,
which are connected by a pass that leads through the saddle point.
An atom at the saddle point is in the transition (activated) state
[4]. Under the action of an applied stress the forward velocity of
a (plastic) flow unit is the net number of times it moves forward,
multiplied by the distance it jumps. Eyring proposed a specific
molecular model of the amorphous structure and a mechanism of
flow. With reference to this idea, this mechanism results in a
(shear) strain rate given by
\begin{equation}  
 \dot{\xi}=2\frac{V_h}{V_m}\frac{k_B T}{h}\exp (\frac{-\delta
G}{k_B T}) \sinh(\frac{V_h \tau}{2 k_B T})
\end{equation}
where
$V_h=\lambda_2\lambda_3\lambda$,$V_m=\lambda_2\lambda_3\lambda_1$,
$\lambda_1$ is the perpendicular distance between two neighboring
layers of molecules sliding past each other, $\lambda$ is the
average distance between equilibrium positions in the direction of
motion, $\lambda_2$ is the distance between neighboring molecules
in this same direction (which may or may not equal $\lambda$),
$\lambda_3$ is the molecule to molecule distance in the plane
normal to the direction of motion, and $\tau$ is the local applied
stress, $\delta G$ is the activation energy, $h$ is the Planck
constant, $k_B$ is the Boltzmann constant, $T$ is the temperature,
$V_h$ is the activation volume for the molecular event [4-5]. The
flow of the
 glass is envisaged as the propagation of kinks
in the molecules into available holes. In order for the motion of
the kink to result in a plastic flow, it must be raised
(energised) into the activated state and pass over the saddle
point. 
\newline
Solving Eqn. (1) for the force or $\tau$, one obtains:
\begin{equation}
  \tau=\frac{2 k_B T}{V_h} \sinh^{-1} (\frac{\dot{\xi}}{B}),
\end{equation}
which in the limit of small $(\dot{\xi}/B)$ reduces to Newton's
law for viscous flow.
\newline
We start to consider a steady, fully developed flow of the glassy
fluid in a wavy-rough microannulus of $r_2$ (in mean-averaged
outer radius) with the outer wall being a fixed wavy-rough surface
: $r=r_2+\epsilon \sin(k \theta)$ and $r_1$ (in mean-averaged
inner radius) with the inner wall being a fixed wavy-rough surface
: $r=r_1+\epsilon \sin(k \theta+\beta)$, where $\epsilon$ is the
amplitude of the (wavy) roughness, and the wave number : $k=2\pi
/L $ ($L$ is the wave length), $\beta$ is the phase shift. The
schematic is illustrated in Fig. 2. \newline Firstly, this glassy
fluid can be expressed as
 $\dot{\xi}=\dot{\xi}_0  \sinh(\tau/\tau_0)$,  $\tau_0 \equiv {2 k_B
 T}/{V_h}$,
where $\dot{\xi}$ is the shear rate, $\tau$ is the shear stress,
and $$\dot{\xi}_0 \equiv B = 2\frac{V_h}{V_m}\frac{k_B T}{h}\exp
(\frac{-\delta G}{k_B T})$$
 is a function of temperature with the dimension of
the shear rate (for small shear stress $\tau \ll \tau_0$,
$\tau_0/\dot{\xi}_0$ represents the viscosity of the material). In
fact, the force balance gives the shear stress at a radius $r$ as
$\tau=-(r \,dp/dz)/2$. $dp/dz$ is the pressure gradient along the
flow (or tube-axis : $z$-axis) direction.\newline Introducing
the forcing parameter 
$\Pi = -(r_2/2\tau_0) dp/dz$
then we have
 $\dot{\xi}= \dot{\xi}_0  \sinh ({\Pi r}/{r_2})$.
As $\dot{\gamma}=- du/dr$ ($u$ is the velocity of the fluid flow
in the longitudinal ($z$-)direction of the microannulus), after
integration, we obtain
\begin{equation}
 u=u_s +\frac{\dot{\xi}_0 r_2}{\Pi} [\cosh \Pi - \cosh (\frac{\Pi r}{r_2})],
\end{equation}
here, $u_s$ is the velocity over the surfaces of the microannulus,
which is determined by the boundary condition. We noticed that
 a general boundary condition for fluid flows over a solid
surface was proposed (cf., e.g.,  [10-11]) as
\begin{equation}
 \delta u=L_s^0 \dot{\xi}
 (1-\frac{\dot{\xi}}{\dot{\xi}_c})^{-1/2},
\end{equation}
where $\delta u$ is the velocity jump over the solid surface,
$L_s^0$ is a constant slip length and $\dot{\xi}_c$ is the
critical shear rate at which the slip length diverges. The value
of $\dot{\xi}_c$ is a function of the corrugation of interfacial
energy. We remind the readers that this expression is based on the
assumption of the shear rate over the solid surface being much
smaller than the critical shear rate of $\dot{\xi}_c$.
$\dot{\xi}_c$ represents the maximum shear rate the fluid can
sustain beyond which there is no additional momentum transfer
between the wall and fluid molecules [10-11]. How generic this
behavior is and whether there exists a comparable scaling for
glassy fluids remain open questions.\newline
With the boundary condition from [10-11], we shall derive the
velocity field and volume flow rate along the wavy-rough
microannuli below using the boundary perturbation technique (cf.,
e.g., [8-9]) and dimensionless analysis. We firstly select
$L_a\equiv r_2 - r_1$ to be the characteristic length scale and
set
$$r'=r/L_a, \hspace*{5mm}
R_o=r_2/L_a, \hspace*{5mm} R_i=r_1/L_a, \hspace*{5mm}
\epsilon'=\epsilon/L_a. $$ After this, for simplicity, we drop all
the primes. It means, now, $r$, $R_o$, $R_i$, and $\epsilon$
become dimensionless. The walls are prescribed as $r=R_o+\epsilon
\sin(k\theta)$, $r=R_i+\epsilon \sin(k \theta+\beta)$ and the
presumed fully-developed flow is along the $z$-direction
(microannulus-axis direction). Along the boundary, we have
 $$\dot{\xi}=({d u}/{d n})|_{{\mbox{\small on surface}}},$$
where, $n$ means the  normal. Let $u$ be expanded in $\epsilon$ :
 $u= u_0 +\epsilon u_1 + \epsilon^2 u_2 + \cdots$,
and on the boundary, we expand $u(r_0+\epsilon dr,
\theta(=\theta_0))$ into
\begin{displaymath}
u(r,\theta) |_{(r_0+\epsilon dr ,\theta_0)}
=u(r_0,\theta)+\epsilon [dr \,u_r (r_0,\theta)]+ \epsilon^2
[\frac{dr^2}{2} u_{rr}(r_0,\theta)]+\cdots=
\end{displaymath}
\begin{equation}
 \hspace*{12mm} \{u_{slip} +\frac{\dot{\xi} R_o}{\Pi} \cosh
 (\frac{\Pi \bar{r}}{R_o})|_r^{R_o+\epsilon \sin(k\theta)},
\end{equation}
where the subscript means the partial differentiation (say, $u_r
\equiv
\partial u/\partial r$) and
\begin{equation}
 u_{slip}|_{{\mbox{\small on surface}}}=L_s^0 \dot{\xi} [(1-\frac{\dot{\xi}}{\dot{\xi}_c})^{-1/2}]
 |_{{\mbox{\small on surface}}},
\end{equation}
\begin{equation}
 u_{{slip}_0}= L_s^0 \dot{\xi}_0 [\sinh\Pi(1-\frac{\dot{\xi}_0 \sinh\Pi}{
 \dot{\xi}_c})^{-1/2}].
\end{equation}
Now, on the outer wall (cf., e.g., [8-9]), 
\begin{displaymath}
\dot{\xi}=\frac{du}{dn}=\nabla u \cdot \frac{\nabla
(r-R_o-\epsilon \sin(k\theta))}{| \nabla (r-R_o-\epsilon
\sin(k\theta)) |}=[1+\epsilon^2 \frac{k^2}{r^2}  \cos^2
(k\theta)]^{-\frac{1}{2}} [u_r |_{(R_o+\epsilon dr,\theta)} -
\end{displaymath}
\begin{displaymath}  
 \hspace*{12mm} \epsilon \frac{k}{r^2}
\cos(k\theta) u_{\theta} |_{(R_o+\epsilon dr,\theta)}
]=u_{0_r}|_{R_o} +\epsilon [u_{1_r}|_{R_o} +u_{0_{rr}}|_{R_o}
\sin(k\theta)-
\end{displaymath}
\begin{displaymath}
  \hspace*{12mm}  \frac{k}{r^2} u_{0_{\theta}}|_{R_o} \cos(k\theta)]+\epsilon^2 [-\frac{1}{2} \frac{k^2}{r^2} \cos^2
(k\theta) u_{0_r}|_{R_o} + u_{2_r}|_{R_o} + u_{1_{rr}}|_{R_o} \sin(k\theta)+ 
\end{displaymath}
\begin{equation}
   \hspace*{12mm} \frac{1}{2} u_{0_{rrr}}|_{R_o} \sin^2 (k\theta) -\frac{k}{r^2}
\cos(k\theta) (u_{1_{\theta}}|_{R_o} + u_{0_{\theta r}}|_{R_o}
\sin(k\theta) )] + O(\epsilon^3 ) .
\end{equation}
Considering $L_s^0 \sim R_o, R_i \gg \epsilon$ case, we presume
$\sinh\Pi \ll \dot{\xi}_c/\dot{\xi_0}$ so that we can
approximately replace
$[1-(\dot{\xi}_0 \sinh\Pi)/\dot{\xi}_c]^{-1/2}$
by
$[1+\dot{\xi}_0 \sinh\Pi/(2 \dot{\xi}_c)]$.
With equations (5), (6) and (8), using the definition of
$\dot{\xi}$, we can derive the velocity field up to the second
order. The key point is to firstly obtain the slip velocity along
the boundary or surface.  \newline After lengthy mathematical
manipulations and using
 $(1-{\dot{\xi}}/{\dot{\xi}_c})^{-1/2}\approx 1+{\dot{\xi}}/({2
 \dot{\xi}_c})$,
\begin{displaymath}
 u_0=-\frac{\dot{\xi}_0 R_o}{\Pi} [\cosh
 (\frac{\Pi r}{R_o})-\cosh \Pi]+u_{{slip}_0},  \hspace*{12mm} u_1=
 \dot{\xi}_0 \sin (k\theta) \sinh \Pi +u_{{slip}_1},
\end{displaymath}
we have %
\begin{displaymath}
 u_{slip}=L_s^0 \{[-u_{0_r}(1-\frac{u_{0_r}}{2
 \dot{\xi}_c})]|_{r=R_o}+\epsilon
 [-u_f(1-\frac{u_{0_r}}{\dot{\xi}_c})]|_{r=R_o}+\epsilon^2
 [\frac{u_f^2}{2\dot{\xi}_c}-u_{sc}
 (1-\frac{u_{0_r}}{\dot{\xi}_c})]|_{r=R_o}\}=
\end{displaymath}
\begin{equation}
 \hspace*{36mm} u_{slip_0} +\epsilon \,u_{slip_1} + \epsilon^2 u_{slip_2}
 +O(\epsilon^3)
\end{equation}
where
\begin{displaymath}
 u_f =u_{1_r} + u_{0_{rr}} \sin (k\theta)-\frac{k}{r^2} \cos (k\theta)
 u_{0_{\theta}}=-\frac{\Pi}{R_o}\dot{\xi}_0 \cosh(\frac{\Pi}{R_o}r) \, \sin (k\theta),
\end{displaymath}
and
\begin{displaymath}
 u_{sc} =-\frac{k^2}{2 r^2}\cos^2 (k \theta) u_{0_r} +\frac{1}{2}
 u_{0_{rrr}} \sin^2 (k\theta)=\frac{1}{2}\dot{\xi}_0 [\frac{k^2}{2 r^2}\cos^2 (k \theta)
 -\frac{\Pi^2}{R_o^2}\sin^2 (k\theta)]\sinh(\frac{\Pi}{R_o}r).
\end{displaymath}
Thus, at $r=R_o$, up to the second order,
\begin{displaymath}
 u_{slip}\equiv u_s=L_s^0 \dot{\xi}_0  \sinh\Pi(1+\frac{K_0}{2})+\epsilon \dot{\xi}_0 \sin(k\theta)
 [\sinh \Pi+ \frac{\Pi}{R_o}L_s^0\cosh\Pi \, (1+K_0)]+\epsilon^2 L_s^0\frac{\dot{\xi}_0 }{2}\{
 [
\end{displaymath}
\begin{equation}
 \frac{\Pi \cosh \Pi}{R_o L_s^0}  \sin^2 (k\theta)-\frac{k^2}{R_o^2} \cos^2 (k\theta)+
 \frac{\Pi^2}{R_o^2} \sin^2 (k\theta)]\sinh\Pi (1+K_0) +
 \frac{\Pi^2}{R_o^2} \frac{\dot{\xi}_0}{\dot{\xi}_c}
  \cosh^2 \Pi \,\sin^2 (k\theta) \},
\end{equation}
where $K_0=1+({\dot{\xi}_0
 \sinh\Pi})/{\dot{\xi}_c}$
From the velocity fields (up to the second order), we can integrate
them with respect to the cross-section to get the volume flow rate
($Q$, also up to the second order here).
%
 $ Q=\int_0^{\theta_p} \int_{R_i+\epsilon \sin(k\theta+\beta)}^{R_o+\epsilon \sin(k\theta)}
 u(r,\theta) r
 dr d\theta =Q_{smooth} +\epsilon\,Q_{p_0}+\epsilon^2\,Q_{p_2}$.
\begin{displaymath}
 Q=\pi \dot{\gamma}_0 \{L_s^0 (R_o^2-R_i^2)  \sinh\Pi \,
 (1-\frac{\sinh\Pi}{\dot{\xi}_c/\dot{\xi_0}})^{-1/2}+
 \frac{R_o}{\Pi}[(R_o^2-R_i^2)\cosh\Pi-\frac{2}{\Pi}(R_o^2 \sinh \Phi-
\end{displaymath}
\begin{displaymath}
  R_i R_o \sinh(\Pi \frac{R_i}{R_o})+ \frac{2 R_o^2}{\Pi^2}(\cosh\Pi-\cosh(\Pi \frac{R_i}{R_o})]\}+
  \epsilon^2 \{\pi\dot{\xi}_0 [\Pi \frac{\cosh \Pi}{4}
  (R_o-\frac{R_i^2}{R_o})]+
\end{displaymath}
\begin{displaymath}
 L_s^0\frac{\pi}{4} \dot{\xi}_0  \sinh \Pi (1+\frac{\sinh\Pi}{\dot{\xi}_c/\dot{\xi_0}})
 (-k^2+\Pi^2)[1-(\frac{R_i}{R_o})^2]+
 \frac{\pi}{2}  [(u_{slip_0} +
\end{displaymath}
\begin{displaymath}
\frac{\dot{\xi}_0\,R_o}{\Pi} \cosh\Pi) +\dot{\xi}_0 R_o(
  -\sinh\Pi +\frac{\cosh\Pi}{\Pi})+ \dot{\xi}_0(R_i \sinh
  (\Pi \frac{R_i}{R_o})-\frac{R_o}{\Pi} \cosh (\Pi
  \frac{R_i}{R_o}))]+
\end{displaymath}
\begin{displaymath}
  \pi \dot{\xi}_0 \{[\sinh\Pi+\Pi\frac{\cosh\Pi}{R_o}
  (1+\frac{\sinh\Pi}{\dot{\xi}_c/\dot{\xi_0}})] [R_o-R_i (\cos\beta
  +\sin\beta)]\}+
\end{displaymath}
\begin{equation}
  L_s^0\frac{\pi}{4} \Pi^2 \dot{\xi}_0 \frac{\cosh\Pi}{\dot{\xi}_c/\dot{\xi}_0}[1
-(\frac{R_i}{R_o})^2
 ]
 \cosh\chi.
\end{equation}
Here,
\begin{equation}
 u_{{slip}_0}= L_s^0 \dot{\gamma}_0 [\sinh\Pi(1-\frac{\sinh\Pi}{
 \dot{\xi}_c/\dot{\xi}_0})^{-1/2}].
\end{equation}
\section{Results and discussion}
In the following, as our interest is about the transport of glassy
matter at very low temperature environment ($\Pi \rightarrow 0$),
thus we should take the asymptotic limit of $Q$. In fact, we have,
for the smooth surface of the micropore,
\begin{displaymath}
 Q_{smooth}|_{\Pi\rightarrow 0}=\pi \dot{\xi}_0 \{L_s^0 (R_o^2-R_i^2)  \sinh\Pi \,
 (1-\frac{\sinh\Pi}{\dot{\xi}_c/\dot{\xi_0}})^{-1/2}+
 \frac{R_o}{\Pi}[(R_o^2-R_i^2)\cosh\Pi-
\end{displaymath}
\begin{equation}
  \hspace*{12mm} \frac{2}{\Pi}(R_o^2 \sinh \Pi-R_i R_o \sinh(\Pi \frac{R_i}{R_o}))+ \frac{2 R_o^2}{\Pi^2}
  (\cosh\Pi-\cosh(\Pi \frac{R_i}{R_o}))]\}|_{\Pi\rightarrow 0}=0,
\end{equation}
and for the wavy-rough surface (up to the second order) $Q_{p2}
=0$ as $\Pi \rightarrow 0$.
Now, the (referenced) shear-rate
\begin{equation}
\dot{\xi}_0 \equiv B = 2\frac{V_h}{V_m}\frac{k_B T}{h}\exp
(\frac{-\delta G}{k_B T}),
\end{equation}
is a  function of temperature and the activation energy. From
equation (11), we can observe that the  transport
 of glassy matter in very-low temperature : $T\rightarrow 0$
 is rather small ($Q_{p2}\propto \dot{\xi}_0\propto T \exp(-1/T)$)! It is worth pointing out that
 the Eyring model requires the
interaction between atoms in the direction perpendicular to the
shearing direction for the momentum transfer  [4-5]. This might
explain why our result is orientation dependent (due to
$\beta$).\newline
Note that, based on the rate-state Eyring model [4] (of
stress-biased thermal activation), structural rearrangement is
associated with a single energy barrier (height) $E$ that is
lowered or raised linearly by an applied stress $\sigma$ or
$\tau$. If the transition rate is proportional to the shear strain
rate (with a constant ratio : $C_0$), we have
\begin{equation}
\sigma = E / V^*+( k_B T / V^*)
 \ln ( \dot{\xi} /C_0 \nu_0),
\end{equation}
where $V^*$ is a constant called the activation volume, $k_B$ is
the Boltzmann constant, $T$ is the temperature, and $\nu_0$ is an
attempt frequency [4-5,12]. Normally, the value of $V^*$ is
associated with a typical volume required for a molecular shear
rearrangement. Thus, the nonzero flow rate (of the glass) as
forcing is absent could be related to a barrier-overcoming or
tunneling  for shear-thinning matter along the wavy-roughness
(geometric valley and peak served as atomic potential surfaces) in
annular micropores when the wavy-roughness is present. Once the
geometry-tuned potentials (energy) overcome this barrier, then the
tunneling (spontaneous transport) inside
wavy-rough annular micropores occurs. \newline 
We also noticed that, as described in [4], mechanical loading
lowers energy barriers, thus facilitating progress over the
barrier by random thermal fluctuations. The simplified Eyring
model approximates the loading dependence of the barrier height as
linear. 
The linear
dependence will always correctly describe small changes in the
barrier height, since it is simply the first term in the Taylor
expansion of the barrier height as a function of load. It is thus
appropriate when the barrier height changes only slightly before
the system escapes the local energy minimum. This situation occurs
at higher temperatures; for example, Newtonian flow is obtained in
the Eyring model in the limit where the system experiences only
small changes in the barrier height before thermally escaping the
energy minimum. As the temperature decreases, larger changes in
the barrier height occur before the system escapes the energy
minimum (giving rise to, for example, non-Newtonian flow). In this
regime, the linear dependence is not necessarily appropriate, and
can lead to inaccurate modelling. This explains why we should
adopt the hyperbolic sine law [4-5] to treat the glassy matter.
\newline
Finally, we present the calculated maximum velocity (unit : m/s)
with respect to the temperature in Fig. 3. Geometric parameters :
$r_2=100$ nm, the activation volume : $0.2$ nm$^3$ and the
roughness amplitude $\epsilon=0.05 r_2$. We consider the effect of
the activation energy : $10^{22}$ and $2\times 10^{22}$ Joule.
Around $T\sim 0.25 ^{\circ}$K, the maximum velocity (of the glassy
matter) either keeps decreasing as the temperature increases for
larger activation energy  or instead increases as the temperature
increases for smaller activation energy! The latter observation
might be related to the argues raised in [13] for annealing
process of solid helium at similar low temperature environment if
we treat the solid helium to be glassy at low temperature regime.
\section{Conclusion}
To conclude in brief, we obtain  the transport of glassy matter
inside annular wavy-rough micropores
under very low temperature environment. 
The flow rate is rather small (of the order of magnitude of the
square of the small wavy-roughness amplitude) and is proportional
to the (referenced) shear rate (which is strongly temperature as
well as activation energy dependent), the slip length and phase
shift of the wavy-roughness as illustrated above. {\it
Acknowledgement.} 
The author stayed at the Chern Shiing-Shen Institute of
Mathematics, Nankai University around the beginning of 2008-Jan.
Thus the author should thank their hospitality for the first stage
Visiting-Scholar Program.
%

\newpage

\psfig{file=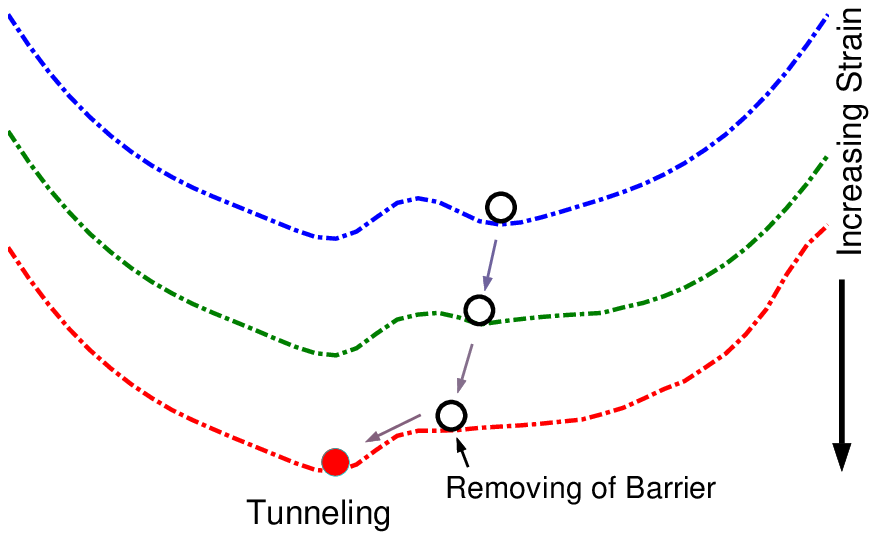,bbllx=-1.0cm,bblly=18.4cm,bburx=10cm,bbury=26.8cm,rheight=8cm,rwidth=8cm,clip=}

\begin{figure}[h]
\hspace*{10mm} Fig. 1.  \hspace*{1mm} The structural contribution
to the shear stress is : Shear thinning. Increasing strain causes
a local energy minimum to flatten until it disappears (removing of
energy barrier or quantum-like tunneling).
\end{figure}

\vspace{24mm}

\psfig{file=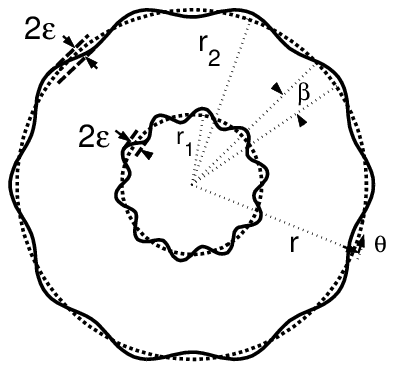,bbllx=-1.0cm,bblly=20.5cm,bburx=8cm,bbury=26.5cm,rheight=4cm,rwidth=7cm,clip=}

\begin{figure}[h]
\hspace*{10mm} Fig. 2. \hspace*{1mm} Schematic  of an annular
micropore. $\beta$ is the phase shift between the outer
\newline \hspace*{10mm} and inner wavy-roughness. $\epsilon$ is
the amplitude of small wavy-roughness.
\end{figure}

\newpage

\vspace{6mm}

\psfig{file=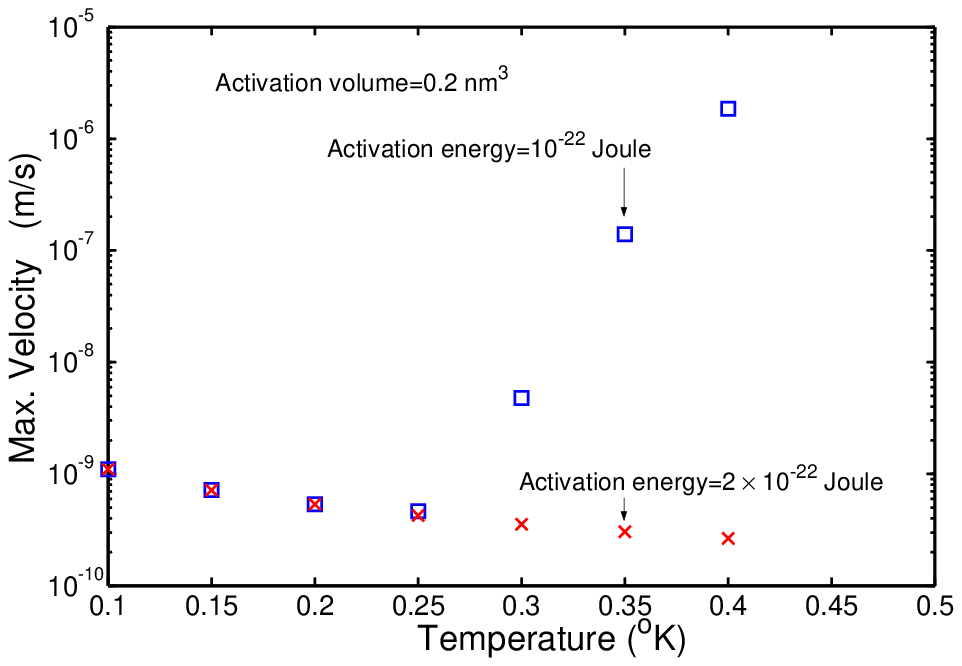,bbllx=-1.0cm,bblly=19.5cm,bburx=12cm,bbury=27cm,rheight=7.5cm,rwidth=9.5cm,clip=}

\begin{figure}[h]
\hspace*{10mm} Fig. 3.  \hspace*{1mm} Comparison of calculated
(maximum) velocity (unit : m/s) using two activation \newline
\hspace*{10mm} energies $10^{-22}$ and $2\times 10^{-22}$
 Joule. Around $T\sim 0.25 ^{\circ}$K, the monotonic trend of velocity
  \newline \hspace*{10mm} bifurcates as the temperature increases.
  $r_2=100$ nm and $\epsilon=0.05 r_2$ here.
\end{figure}

\end{document}